\documentclass[a4paper,fleqn,usenatbib]{mnras}

\usepackage[T1]{fontenc}
\usepackage{ae,aecompl}

\usepackage{graphicx}	
\usepackage{amsmath}	
\usepackage{amssymb}	
\usepackage{subfigure}

\newcommand{\dinamo}{{\sc dinamo}}

\newcommand{\csg}{\,{\rm cm^2 \, g^{-1}}}
\newcommand{\um}{\, {\rm \mu m}}

\newcommand{\kel}{\, {\rm K}}

\newcommand{\msun}{\, {\rm M}_\odot}

\newcommand{\pc}{\, {\rm pc}}

\newcommand{\mpc}{\, {\rm Mpc}}

\newcommand{\kl}{\kappa_{\lambda}}
\newcommand{\kz}{\kappa_0}
\newcommand{\kfh}{\kappa_{500}}
\newcommand{\sism}{\Sigma_{\rm ISM}}
\newcommand{\av}{A_{\rm V}}

\newcommand{\fext}{f_{\rm att}}

\title[The mass opacity of interstellar dust]{The apparent anti-correlation between the mass opacity of interstellar dust and the surface-density of interstellar gas}

\author[Priestley \& Whitworth]{
F. D. Priestley and A. P. Whitworth
\\
School of Physics and Astronomy, Cardiff University, Queen's Buildings, The Parade, Cardiff CF24 3AA, UK \\
}

\date{Accepted XXX. Received YYY; in original form ZZZ}

\pubyear{2020}

\begin{document}
\label{firstpage}
\pagerange{\pageref{firstpage}--\pageref{lastpage}}
\maketitle

\begin{abstract}

Recent analyses of {\it Herschel} observations suggest that in nearby disc galaxies the dust mass opacity at $500 \um$, $\kfh$, decreases with increasing gas surface density, $\sism$ \citep{clark2019}. This apparent anti-correlation between $\kfh$ and $\sism$ is opposite to the behaviour expected from theoretical dust evolution models; in such models, dust in denser, cooler regions (i.e. regions of increased $\sism$) tends to grow and therefore to have increased $\kfh$. We show, using a toy model, that the presence of a range of dust temperatures along the line of sight can lead to spuriously low estimated values of $\kfh$. If in regions of higher $\sism$ the range of dust temperatures extends to lower values (as seems likely), the magnitude of this effect may be sufficient to explain the apparent anti-correlation between $\kfh$ and $\sism$. Therefore there may not be any need for spatial variation in the intrinsic dust properties that run counter to theoretical expectations.

\end{abstract}



\section{Introduction}

Dust is an important constituent of the interstellar medium (ISM), making up $\sim\!1\%$ of the total mass, and locking up around half of the available metals \citep{draine2011}. Far-infrared (IR) observations of thermal dust emission are an important tool in investigating the properties of both the dust and the ISM at large \citep{kennicutt2009,eales2012}. However, converting observed fluxes into physical quantities requires knowledge of the optical properties of interstellar dust at far-infrared and sub-millimetre wavelengths. A commonly used simplification is to fit the spectral energy distribution (SED) with a modified blackbody (MBB) function,
\begin{eqnarray}
\label{eq:mbb}
S_{\lambda} &=& \frac{M_{\rm d}\,\kl\,B_{\lambda}\!(T)}{D^2}\,,
\end{eqnarray}
where $S_{\lambda}$ is the observed flux density, $M_{\rm d}$ is the dust mass, $\kl$ is the dust mass opacity (i.e. the opacity per unit mass of dust)  at wavelength $\lambda$, $B_{\lambda}(T)$ is the Planck Function (i.e. the blackbody intensity), $T$ is the dust temperature, and $D$ is the source distance. Eqn. (\ref{eq:mbb}) is only accurate if all the dust on a given line of sight (a) is in thermal equilibrium, (b) has the same temperature, { and (c) is optically thin to its own emission (generally true in the far-IR)}. $\kl$ can be measured in the laboratory, or calculated (if the dust composition and the distributions of size and shape are specified). The variation of $\kappa_\lambda$ with wavelength can usually be approximated with a power law in the far-infrared and sub-millimetre { (e.g. Figure 4 from \citealt{galliano2018})}, viz.
\begin{eqnarray}
  \label{eq:beta}
  \kl &\simeq& \kappa_0 \left(\frac{\lambda_0}{\lambda}\right)^{\beta}\,;
\end{eqnarray}
here $\kappa_0$ is the mass opacity at the reference wavelength $\lambda_0$. If $\kappa_0$ and $D$ are known, $M_{\rm d}$, $T$ and $\beta$ can -- in principle -- be fit as free parameters.

The value of $\kz$ used can be based on laboratory measurements (e.g. \citealt{jaeger1994,zubko1996}) or on theoretical calculations for a particular material \citep[e.g.][]{laor1993}, but usually an average value is adopted, accounting for a range of dust compositions and sizes, { constrained by observational features such as the extinction curve and elemental depletions}, and in some cases also { taking account of} the evolutionary processes undergone by dust in different phases of the ISM (e.g. \citealt{draine2007,jones2016}). Derived dust properties are therefore very model dependent, and although most commonly-used values for $\kz$ agree within a factor of order six, the full range is at least a factor of 300 \citep[see Table D1 in][]{Whitworthetal2019}. Moreover, there is no guarantee that any of the models invoked accurately { reproduce the true properties of} interstellar dust, or that the same { model} is appropriate in all environments  (i.e. different phases of the ISM, galaxies with different stellar and/or gas properties).

\citet{james2002} { have} proposed a way to estimate $\kz$ in galaxies directly, by determining the mass and metallicity of the interstellar gas, and assuming a constant fraction of the metallicity is locked up in dust; this gives a value for the dust mass independent of the far-infrared data. The fraction of the metallicity locked up in dust grains ($0.5\pm 0.1$) appears to be relatively constant in the local Universe \citep[e.g.][]{jenkins2009,peeples2014}, and so the errors introduced by this assumption are probably lower than those affecting theoretical estimates for $\kz$. The measured value of $\kz$ can then be used to analyse dust in other galaxies, and to calibrate theoretical dust models. \citet{clark2016} present an updated estimate of $\kz$ obtained by this procedure, taking advantage of improved far-infrared observations from {\it Herschel}. They find $\kfh \sim 0.5\,{\rm cm^2\,g^{-1}}$ (where $\kfh$ is $\kl$ at $\lambda\!=\!500 \um$). This is somewhat smaller than the values { predicted by} recent theoretical models \citep[e.g.][]{jones2016}, but consistent with some earlier models \citep[e.g.][]{draine2003}.

Recently, \citet{clark2019} have applied this method to spatially resolved observations of { the} two galaxies { M74 (NGC628) and M83 (NGC 5236);} they conclude that $\kfh$ { tends to decrease} with increasing ISM surface density, $\sism$. This is in contrast with what is expected from dust evolution models, where grains in denser regions of the ISM are expected to be larger and therefore to have higher far-IR opacities \citep[e.g.][]{kohler2015,ysard2018}. \citet{clark2019} find this apparent anti-correlation between $\kfh$ and $\sism$ to be robust against systematic effects that might have affected their methodology. { \citet{bianchi2019} analysed a large sample of galaxies and found that the dust emissivity at $250 \um$ appears to decrease with increasing molecular gas fraction. If interpreted as a decline in the opacity, this would appear to support the $\kfh$-$\sism$ anti-correlation.} However, the calculation of $\kfh$ depends on the single dust temperature { returned} by a single-temperature MBB fit { (hereafter, a 1MBB fit)}. If there is a range of temperatures on the line of sight, this estimate, being flux-weighted, is biased towards the higher temperatures. The procedure used by \citet{clark2019} then tends to underestimate $\kfh$ in order to compensate for this temperature bias. In this paper, we show that, with reasonable physical assumptions, { this} temperature bias can produce an apparent anti-correlation between $\kfh$ and $\sism$, similar in magnitude to that found by \citet{clark2019}, without any underlying variation in the intrinsic optical properties of the dust.

\section{Method}

\begin{figure}
\centering
\includegraphics[width=\columnwidth]{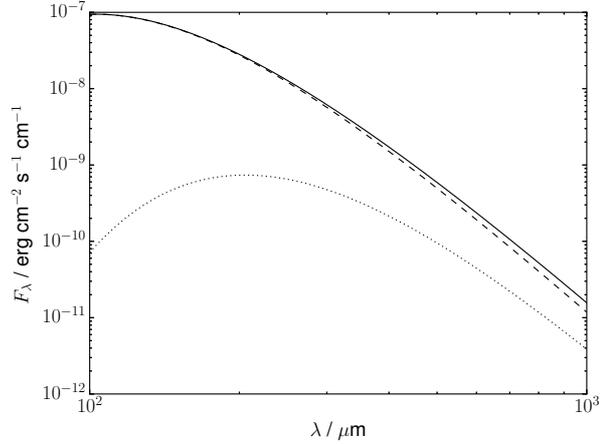}
\caption{The fluxes from MBBs { with identical mass} of $T\!=\!20 \kel$ (dashed line) and $T\!=\!10 \kel$ (dotted line), and their sum (solid line). Even at long wavelengths the $10\,{\rm K}$ contribution is small.}
\label{fig:bbfit}
\end{figure}

\citet{clark2019} estimate $\kfh$ by rearranging Equation (\ref{eq:mbb}) so that
\begin{equation}
  \label{eq:kappa}
  \kfh = \frac{D^2\,S_{500}}{M_{\rm d}\,B_{500}(T)}
\end{equation}
with $M_{\rm d}$ calculated from the gas mass, { the} metallicity and { the assumed fraction of the metallicity locked up in dust grains}. The dust temperature is estimated by fitting a single-temperature MBB to the {\it Herschel} far-IR fluxes. As the emitted flux increases with temperature at all wavelengths, a given mass of warmer dust will contribute more to the total emission than the same mass of cooler dust, particularly at shorter wavelengths, thereby biasing the { values} returned by the MBB fit \citep{shetty2009b}. As a simple illustration of this effect, Figure \ref{fig:bbfit} shows the monochromatic { fluxes} from two MBBs, calculated using Equations (\ref{eq:mbb}) and (\ref{eq:beta}), and { the sum of these fluxes}. We choose typical (but arbitrary) values of $M_{\rm d} = 10^5 \msun$, $D = 1 \mpc$, $\beta = 2$ and $\kfh = 2 \csg$. Fitting a single MBB to the { sum of the fluxes} gives a temperature of $21 \kel$, and a slightly reduced $\beta=1.8$. Substituting the total mass of $2 \times 10^5 \msun$ into Equation (\ref{eq:kappa}) then returns a value of $\kfh = 1.1 \csg$, nearly a factor of 2 smaller than the true value.

\citet{clark2019} acknowledge this issue, but argue that this will only systematically lower their measured $\kfh$ from the true value. They also note that repeating their analysis using a two{ -temperature} MBB fit { (hereafter a 2MBB fit)}, 
does not remove the anti-correlation between $\kfh$ and $\sism$. However, they do not appear to have considered the possibility that the temperature bias might be correlated with the ISM surface density, $\sism$, in the sense that at larger column density, attenuation of the radiation field by dust extinction is likely to be greater, leading to both lower dust temperatures and/or a higher proportion of dust at lower temperatures.

We investigate this effect by using a toy model to generate synthetic dust SEDs, { and then using the same procedure as \citet{clark2019} to estimate $\kfh$ from these synthetic dust SEDs.} { The toy model has two parameters, $\fext$ and $\sism$. A fraction $(1-\fext)$ of the modelled dust is heated by an unattenuated radiation field, and the remainder, $\fext$, by an attenuated radiation field. The attenuation is related to $\sism$ using the standard ratio between visual extinction, $\av$, and the column-density of hydrogen in all forms, $N_{_{\rm H}}$, \citep{bohlin1978} viz.}
\begin{eqnarray}\nonumber
\av\!\!&\!\!\simeq\!\!&\!\!0.6\,{\rm mag}\left(\!\frac{N_{_{\rm H}}}{10^{21}\,{\rm H\,cm^{-2}}}\!\right)\,\simeq\,0.052\,{\rm mag}\left(\!\frac{\sism}{\msun\,{\rm pc}^{-2}}\!\right)\!.\\\label{EQN:AV2SISM}
\end{eqnarray}
Although $\fext$ is probably correlated with $\sism$, we treat $\fext$ and $\sism$ as independent free parameters, so as  to avoid additional assumptions.

The synthetic SEDs are generated using the dust heating code \dinamo{} \citep{priestley2019}, { which} accounts for stochastic heating of small grains. We assume that dust is heated by the \citet{mathis1983} interstellar radiation field, { and attenuation is implemented} using the \citet{cardelli1989} extinction law. We adopt a power law distribution of grain sizes between $0.005\;\mbox{and}\;0.25 \um$, with exponent $-3.5$ \citep{mathis1977}. { We compute the dust properties using} optical constants for amorphous carbon from \citet{zubko1996}. { We address the possible evolution of the dust properties with $\sism$ in Appendix \ref{sec:change}.} { We emphasise that, since the temperature bias is a differential effect, the results we present below are not sensitive to these choices of radiation field, grain size distribution, or grain material.} 

{ We convert the dust SED into photometric fluxes for the PACS and SPIRE bands using the appropriate filter response curves, and assume error bars of $10\%$ for each band, typical of the average values of the DustPedia sample \citep{davies2017} from which \citet{clark2019} take their far-IR data.} We can then derive the value of $\kfh$ that would be inferred using the \citet{clark2019} procedure, as a function of $\sism$, for representative values of $\fext$. For the purpose of illustration, we again assume a source distance of $D = 1 \mpc$ and a total dust mass $M_{\rm d} = 10^5 \msun$. However, since the dependence on these parameters is limited to $S_\lambda\propto M_{\rm d}/D^2$ { (see Eqn. \ref{eq:kappa})}, this does not affect the calculation of $\kfh$.

We note that there are two distinct effects producing a range of temperatures. First, even if the radiation field heating the dust were the same everywhere, dust grains of different composition and/or different size would have different temperatures; the smallest dust grains would also have time-varying temperatures. Second, the radiation field heating the dust is not the same everywhere (this is what we parameterise, very simplistically, with $\fext$), and therefore { even large} dust grains of the same composition and size have different temperatures in different locations.

\section{Results}

\begin{figure}
\centering
\includegraphics[width=\columnwidth]{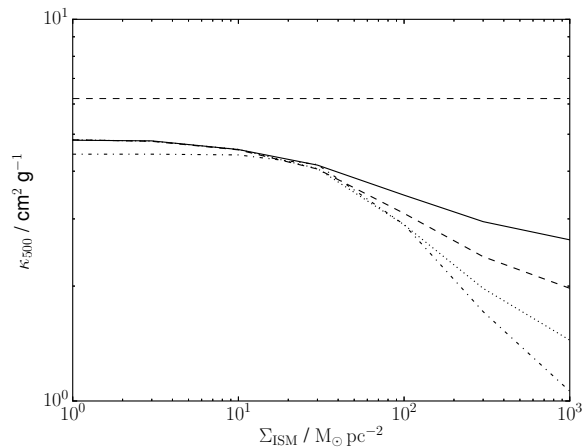}
\caption{The inferred value of $\kfh$ as a function of $\av$, obtained using a single-temperature MBB fit. The extinction, $\av$, which determines the attenuation of the radiation field, is related to the surface-density, $\sism$, by Eqn. (\ref{EQN:AV2SISM}). The different curves represent cases where different fractions of dust, $\fext$, are located in regions where the radiation field is attenuated: $\fext = 0.5$ (solid line), $0.667$ (dashed line), $0.75$ (dotted line) and $0.9$ (dot-dashed line). The thin dashed line shows the true value of $\kfh = 6.2 \csg$.}
\label{fig:ks}
\end{figure}

Figure \ref{fig:ks} shows how the estimated $\kfh$ varies with $\sism$, for different discrete values of $\fext$ between $0.5$ and $0.9$. { For clarity, we do not show error bars, but uncertainties are $10\%$ or lower, for all combinations of $\fext$ and $\sism$.} The inferred $\kfh$ { values are} always below the true value ($6.2 \csg$), even { when there is} no attenuation, { because a 1MBB} fit cannot account for multiple grain sizes, each with their own temperature (or temperatures in the case of very small stochastically heated grains). { Using a 1MBB} fit (Figure \ref{fig:ks}), { the derived} $\kfh$ decreases monotonically with $\sism$ for all values of $\fext$, and by a factor between 2 and 5 over two orders of magnitude in $\sism$, depending on the value of $\fext$. Qualitatively this reproduces the trend seen by \citet{clark2019}, without any change in the underlying dust optical properties.

\begin{figure}
\centering
\includegraphics[width=\columnwidth]{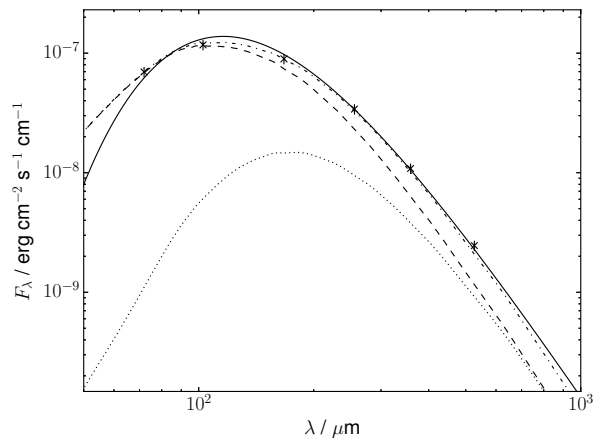}
\caption{{ Fluxes of the unattenuated (dashed line) and attenuated (dotted line) dust components for $\fext = 0.667$ and $\sism = 100 \msun\,{\rm pc}^{-2}$ (hence $\av =5.2\,{\rm mag}$), their combined flux (dot-dashed line) and the best-fit MBB with $T = 20.7 \kel$ and $\beta = 1.0$ (solid line). The photometric fluxes and error bars are marked as crosses.}}
\label{fig:dustfit}
\end{figure}

\begin{figure}
\centering
\includegraphics[width=\columnwidth]{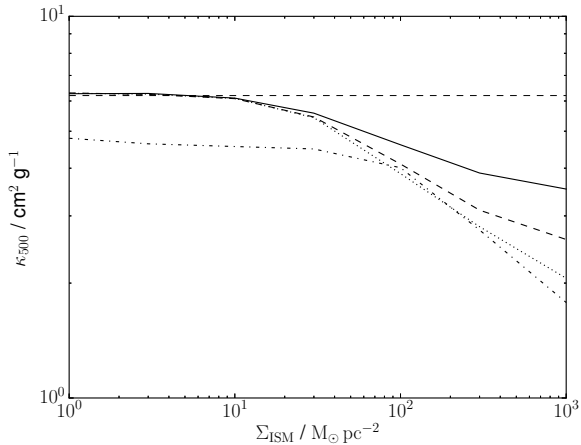}
\caption{{ As Figure \ref{fig:ks}, but for a radiation field strength increased by a factor of $10$.}}
\label{fig:g10}
\end{figure}

The decrease is entirely due to fitting with single-temperature MBBs, SEDs that have been generated { from a wide range of dust temperatures. This is shown in Figure \ref{fig:dustfit}, where the bias towards the temperature of the dust heated by the unattenuated flux is apparent. Temperature values from the 1MBB fits range between $18\kel$ and $22 \kel$, and all fits return $\beta\simeq 1.0$. The dust opacity we use actually has $\beta\simeq 1.4$, but fixing $\beta$ to this value does not significantly change the results; the $\kfh$ values returned by the 1MBB fits are slightly higher, but they still decrease with increasing $\sism$, and by essentially the same factor over the same range of $\sism$.} { Increasing the radiation field strength by a factor of $10$ also fails to eliminate this trend, although in this case, shown in Figure \ref{fig:g10}, the inferred values of $\kfh$ at lower $\sism$ and lower $\fext$ are in better agreement with the true value.}

Quantitatively, the opacities estimated by \citet{clark2019} decrease with increasing surface density, $\sism$, somewhat faster and at somewhat lower values of $\sism$ than in our toy model. Specifically, for $\sism$ between $10$ and $100 \msun \, {\rm pc^{-2}}$, where the majority of the data in \citet{clark2019} lie, we find a power law exponent, $d\ln(\kfh)/d\ln (\sism)$, between $-0.14$ and $-0.28$, whereas \citet{clark2019} find $d\ln(\kfh)/d\ln (\sism)\sim -0.35$ for both the galaxies they analyse (M74 and M83). However, there are two factors which in a more sophisticated model would reduce this discrepancy, by making the model slope steeper (i.e. more negative). First, $\fext$ is very likely correlated with $\sism$. Second, the interstellar medium has substructure on scales below those resolved by the observations ($\sim 100\,{\rm pc}$); consequently the radiation field reaching the dust in these unresolved substructures is more strongly attenuated than the model predicts, and the dust is cooler. If the model were able to take these two factors into account, the slope predicted by the model would be steeper, and the decrease in $\kfh$ would occur at lower values of $\sism$. Given the significant error bars on the observed values, and the simplicity of our model, we conclude that the apparent anti-correlation between $\kfh$ and $\sism$ might be due partly, or even entirely, to the temperature bias being stronger in regions of high $\sism$.

\begin{figure}
\centering
\includegraphics[width=\columnwidth]{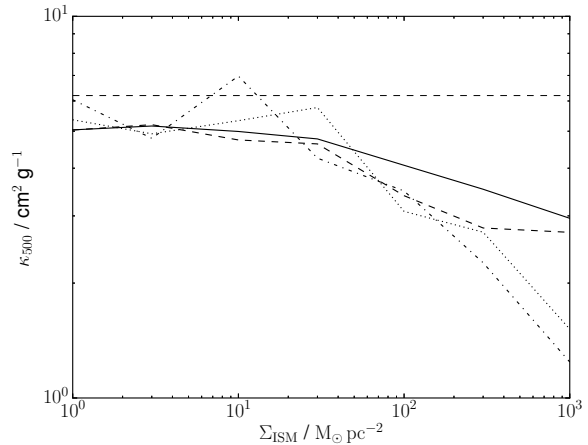}
\caption{As Figure \ref{fig:ks}, but for a two-temperature MBB fit.}
\label{fig:2mbb}
\end{figure}

\citet{clark2019} find that, if they repeat their analysis using a { 2MBB} fit, there is still an apparent anti-correlation between the inferred $\kfh$ and $\sism$, and we find the same. Figure \ref{fig:2mbb} shows the results we obtain with a two-temperature fit. The second MBB results in a slightly higher inferred $\kfh$, closer to the true value, but the anticorrelation with $\sism$ remains. { The uncertainties in $\kfh$ are also much larger than for the single-$T$ fit ($\sim 40\%$).} Although in principle the { 2MBB fit} might be expected to account for the colder dust missed by the { 1MBB} fit, there is no guarantee that a { 2MBB} fit will return mass fractions and temperatures even approximately representative of the true values. In addition (and as also noted by \citealt{clark2019}), using two MBBs with six free parameters overfits the data when only five points (two PACS and three SPIRE bands) are available; this is the cause of the non-monotonic trends on Figure \ref{fig:2mbb}.

\section{Discussion}

We have shown that a relatively simple physical model can produce a spurious apparent anti-correlation between $\kfh$ and $\sism$, similar in magnitude to the one reported by  \citet{clark2019}, when analysed with the same procedure they use. The critical feature of the model is that on any line of sight there is a range of dust temperatures, and this range extends to lower temperatures on lines of sight with higher surface density (and hence, implicitly, higher attenuation of the radiation field heating some of the dust). Whether this is the correct explanation for the apparent anti-correlation, rather than a spatial variation in the intrinsic optical properties of the dust, or some combination of the two, is unclear. The simplistic model explored here assumes that, on any line of sight, a fraction $(1-\fext)$ of the dust grains is exposed to an unattenuated radiation field, and a fraction $\fext$ is exposed to a radiation field attenuated by an $\av$ which is proportional to the surface-density of the interstellar medium, $\sism$, on that line of sight (see Eqn. \ref{EQN:AV2SISM}). There are two effects that the simple model does not include, but which would probably improve correspondence between the model predictions and the observations.

First, it seem likely that the fraction of dust heated by an attenuated radiation field, $\fext$, increases with the surface-density of the interstellar gas, $\sism$; in other words $\fext$ is positively correlated with $\sism$. This would generate a single plot of $\kfh$ against $\sism$ on Figures \ref{fig:ks} and \ref{fig:2mbb}, with a steeper (negative) slope. Provided it did not produce too large an effect, this would bring the slope of the plot closer to that determined by \citet{clark2019}.

Second, the pixel size used by \citet{clark2019} corresponds to a few $100 \pc$ at the distances of the galaxies studied. On these scales, we expect each pixel to contain significant unresolved substructure (e.g. \citealt{elmegreen2002}). If a significant fraction of the dust is contained in unresolved substructures with low volume-filling factor but high density, this dust experiences a much more strongly attenuated radiation field than { our toy} model delivers (the model only invokes an attenuation corresponding to the surface-density averaged over a pixel). Consequently the coefficient on the righthand side of Eqn. (\ref{EQN:AV2SISM}) should be increased, and the distribution of dust temperatures should extend to even lower values, thereby increasing the temperature bias. This will shift the the curves on Figures \ref{fig:ks}, { \ref{fig:g10} and} \ref{fig:2mbb} to smaller values of $\sism$. Again, provided it did not produce too large an effect, this would bring the plot closer to that observed by \citet{clark2019}.

Although we adopt the \citet{mathis1983} radiation field, which is designed to reproduce conditions in the solar vicinity, the effect we have modelled is a differential one, and therefore we can expect similar results in regions where the radiation field is different, { as indicated by the similar trends obtained with an enhanced radiation field (see Fig. \ref{fig:g10}).} A more compelling evaluation of the effect we have considered would require more detailed and realistic modelling, including variations in the ambient unattenuated radiation field and 3D radiative transfer for a distribution of stellar radiation sources and dust \citep[e.g.][]{draine2014,delooze2014,williams2019}. { We have also used constant dust optical properties, whereas in reality these would be expected to vary with $\sism$. In Appendix \ref{sec:change}, we demonstrate that the effect on the inferred value of $\kfh$ is strongly dependent on the assumed variation. Therefore it is not possible to draw any robust conclusions about the behaviour of the dust opacity using the model developed here.} However, it is clear from the results we report here that temperature bias must play a critical role in the interpretation of far-infrared and sub-millimetre dust-continuum observations like those from {\it Herschel}. The apparent anti-correlation between $\kfh$ and $\sism$ may be an entirely spurious consequence of this bias.

\section*{Acknowledgements}

FDP and APW gratefully acknowledge the support of a Consolidated Grant (ST/K00926/1) from the UK Science and Technology Facilities Council (STFC). { We also thank the referee, Simone Bianchi, for his comments, which helped us to improve the original version of this paper.}




\bibliographystyle{mnras}
\bibliography{kappa}

\begin{thebibliography}{}
\makeatletter
\relax
\def\mn@urlcharsother{\let\do\@makeother \do\$\do\&\do\#\do\^\do\_\do\%\do\~}
\def\mn@doi{\begingroup\mn@urlcharsother \@ifnextchar [ {\mn@doi@}
  {\mn@doi@[]}}
\def\mn@doi@[#1]#2{\def\@tempa{#1}\ifx\@tempa\@empty \href
  {http://dx.doi.org/#2} {doi:#2}\else \href {http://dx.doi.org/#2} {#1}\fi
  \endgroup}
\def\mn@eprint#1#2{\mn@eprint@#1:#2::\@nil}
\def\mn@eprint@arXiv#1{\href {http://arxiv.org/abs/#1} {{\tt arXiv:#1}}}
\def\mn@eprint@dblp#1{\href {http://dblp.uni-trier.de/rec/bibtex/#1.xml}
  {dblp:#1}}
\def\mn@eprint@#1:#2:#3:#4\@nil{\def\@tempa {#1}\def\@tempb {#2}\def\@tempc
  {#3}\ifx \@tempc \@empty \let \@tempc \@tempb \let \@tempb \@tempa \fi \ifx
  \@tempb \@empty \def\@tempb {arXiv}\fi \@ifundefined
  {mn@eprint@\@tempb}{\@tempb:\@tempc}{\expandafter \expandafter \csname
  mn@eprint@\@tempb\endcsname \expandafter{\@tempc}}}

\bibitem[\protect\citeauthoryear{{Bianchi} et~al.,}{{Bianchi}
  et~al.}{2019}]{bianchi2019}
{Bianchi} S.,  et~al., 2019, arXiv e-prints, \href
  {https://ui.adsabs.harvard.edu/abs/2019arXiv190912692B} {p. arXiv:1909.12692}

\bibitem[\protect\citeauthoryear{{Bohlin}, {Savage}  \& {Drake}}{{Bohlin}
  et~al.}{1978}]{bohlin1978}
{Bohlin} R.~C.,  {Savage} B.~D.,   {Drake} J.~F.,  1978, \mn@doi [\apj]
  {10.1086/156357}, \href
  {https://ui.adsabs.harvard.edu/abs/1978ApJ...224..132B} {224, 132}

\bibitem[\protect\citeauthoryear{{Cardelli}, {Clayton}  \& {Mathis}}{{Cardelli}
  et~al.}{1989}]{cardelli1989}
{Cardelli} J.~A.,  {Clayton} G.~C.,   {Mathis} J.~S.,  1989, \mn@doi [\apj]
  {10.1086/167900}, \href
  {https://ui.adsabs.harvard.edu/abs/1989ApJ...345..245C} {345, 245}

\bibitem[\protect\citeauthoryear{{Clark}, {Schofield}, {Gomez}  \&
  {Davies}}{{Clark} et~al.}{2016}]{clark2016}
{Clark} C. J.~R.,  {Schofield} S.~P.,  {Gomez} H.~L.,   {Davies} J.~I.,  2016,
  \mn@doi [\mnras] {10.1093/mnras/stw647}, \href
  {https://ui.adsabs.harvard.edu/abs/2016MNRAS.459.1646C} {459, 1646}

\bibitem[\protect\citeauthoryear{{Clark} et~al.,}{{Clark}
  et~al.}{2019}]{clark2019}
{Clark} C.~J.~R.,  et~al., 2019, \mn@doi [\mnras] {10.1093/mnras/stz2257},
  \href {https://ui.adsabs.harvard.edu/abs/2019MNRAS.489.5256C} {489, 5256}

\bibitem[\protect\citeauthoryear{{Davies} et~al.,}{{Davies}
  et~al.}{2017}]{davies2017}
{Davies} J.~I.,  et~al., 2017, \mn@doi [\pasp]
  {10.1088/1538-3873/129/974/044102}, \href
  {https://ui.adsabs.harvard.edu/abs/2017PASP..129d4102D} {129, 044102}

\bibitem[\protect\citeauthoryear{{De Looze} et~al.,}{{De Looze}
  et~al.}{2014}]{delooze2014}
{De Looze} I.,  et~al., 2014, \mn@doi [\aap] {10.1051/0004-6361/201424747},
  \href {https://ui.adsabs.harvard.edu/abs/2014A&A...571A..69D} {571, A69}

\bibitem[\protect\citeauthoryear{{Draine}}{{Draine}}{2003}]{draine2003}
{Draine} B.~T.,  2003, \mn@doi [\araa]
  {10.1146/annurev.astro.41.011802.094840}, \href
  {https://ui.adsabs.harvard.edu/abs/2003ARA&A..41..241D} {41, 241}

\bibitem[\protect\citeauthoryear{{Draine}}{{Draine}}{2011}]{draine2011}
{Draine} B.~T.,  2011, {Physics of the Interstellar and Intergalactic Medium}

\bibitem[\protect\citeauthoryear{{Draine} \& {Li}}{{Draine} \&
  {Li}}{2007}]{draine2007}
{Draine} B.~T.,  {Li} A.,  2007, \mn@doi [\apj] {10.1086/511055}, \href
  {https://ui.adsabs.harvard.edu/abs/2007ApJ...657..810D} {657, 810}

\bibitem[\protect\citeauthoryear{{Draine} et~al.,}{{Draine}
  et~al.}{2014}]{draine2014}
{Draine} B.~T.,  et~al., 2014, \mn@doi [\apj] {10.1088/0004-637X/780/2/172},
  \href {https://ui.adsabs.harvard.edu/abs/2014ApJ...780..172D} {780, 172}

\bibitem[\protect\citeauthoryear{{Eales} et~al.,}{{Eales}
  et~al.}{2012}]{eales2012}
{Eales} S.,  et~al., 2012, \mn@doi [\apj] {10.1088/0004-637X/761/2/168}, \href
  {https://ui.adsabs.harvard.edu/abs/2012ApJ...761..168E} {761, 168}

\bibitem[\protect\citeauthoryear{{Elmegreen}}{{Elmegreen}}{2002}]{elmegreen2002}
{Elmegreen} B.~G.,  2002, \mn@doi [\apj] {10.1086/324384}, \href
  {https://ui.adsabs.harvard.edu/abs/2002ApJ...564..773E} {564, 773}

\bibitem[\protect\citeauthoryear{{Galliano}, {Galametz}  \& {Jones}}{{Galliano}
  et~al.}{2018}]{galliano2018}
{Galliano} F.,  {Galametz} M.,   {Jones} A.~P.,  2018, \mn@doi [\araa]
  {10.1146/annurev-astro-081817-051900}, \href
  {https://ui.adsabs.harvard.edu/abs/2018ARA&A..56..673G} {56, 673}

\bibitem[\protect\citeauthoryear{{Jaeger}, {Mutschke}, {Begemann}, {Dorschner}
  \& {Henning}}{{Jaeger} et~al.}{1994}]{jaeger1994}
{Jaeger} C.,  {Mutschke} H.,  {Begemann} B.,  {Dorschner} J.,   {Henning} T.,
  1994, \aap, \href {http://adsabs.harvard.edu/abs/1994A%26A...292..641J} {292,
  641}

\bibitem[\protect\citeauthoryear{{James}, {Dunne}, {Eales}  \&
  {Edmunds}}{{James} et~al.}{2002}]{james2002}
{James} A.,  {Dunne} L.,  {Eales} S.,   {Edmunds} M.~G.,  2002, \mn@doi
  [\mnras] {10.1046/j.1365-8711.2002.05660.x}, \href
  {https://ui.adsabs.harvard.edu/abs/2002MNRAS.335..753J} {335, 753}

\bibitem[\protect\citeauthoryear{{Jenkins}}{{Jenkins}}{2009}]{jenkins2009}
{Jenkins} E.~B.,  2009, \mn@doi [\apj] {10.1088/0004-637X/700/2/1299}, \href
  {https://ui.adsabs.harvard.edu/abs/2009ApJ...700.1299J} {700, 1299}

\bibitem[\protect\citeauthoryear{{Jones}, {K{\"o}hler}, {Ysard}, {Dartois},
  {Godard}  \& {Gavilan}}{{Jones} et~al.}{2016}]{jones2016}
{Jones} A.~P.,  {K{\"o}hler} M.,  {Ysard} N.,  {Dartois} E.,  {Godard} M.,
  {Gavilan} L.,  2016, \mn@doi [\aap] {10.1051/0004-6361/201527488}, \href
  {https://ui.adsabs.harvard.edu/abs/2016A&A...588A..43J} {588, A43}

\bibitem[\protect\citeauthoryear{{Kennicutt} Robert~C. et~al.,}{{Kennicutt}
  et~al.}{2009}]{kennicutt2009}
{Kennicutt} Robert~C. J.,  et~al., 2009, \mn@doi [\apj]
  {10.1088/0004-637X/703/2/1672}, \href
  {https://ui.adsabs.harvard.edu/abs/2009ApJ...703.1672K} {703, 1672}

\bibitem[\protect\citeauthoryear{{K{\"o}hler}, {Ysard}  \&
  {Jones}}{{K{\"o}hler} et~al.}{2015}]{kohler2015}
{K{\"o}hler} M.,  {Ysard} N.,   {Jones} A.~P.,  2015, \mn@doi [\aap]
  {10.1051/0004-6361/201525646}, \href
  {https://ui.adsabs.harvard.edu/abs/2015A&A...579A..15K} {579, A15}

\bibitem[\protect\citeauthoryear{{Laor} \& {Draine}}{{Laor} \&
  {Draine}}{1993}]{laor1993}
{Laor} A.,  {Draine} B.~T.,  1993, \mn@doi [\apj] {10.1086/172149}, \href
  {http://adsabs.harvard.edu/abs/1993ApJ...402..441L} {402, 441}

\bibitem[\protect\citeauthoryear{{Mathis}, {Rumpl}  \& {Nordsieck}}{{Mathis}
  et~al.}{1977}]{mathis1977}
{Mathis} J.~S.,  {Rumpl} W.,   {Nordsieck} K.~H.,  1977, \mn@doi [\apj]
  {10.1086/155591}, \href
  {https://ui.adsabs.harvard.edu/abs/1977ApJ...217..425M} {217, 425}

\bibitem[\protect\citeauthoryear{{Mathis}, {Mezger}  \& {Panagia}}{{Mathis}
  et~al.}{1983}]{mathis1983}
{Mathis} J.~S.,  {Mezger} P.~G.,   {Panagia} N.,  1983, \aap, \href
  {https://ui.adsabs.harvard.edu/abs/1983A&A...128..212M} {500, 259}

\bibitem[\protect\citeauthoryear{{Peeples}, {Werk}, {Tumlinson}, {Oppenheimer},
  {Prochaska}, {Katz}  \& {Weinberg}}{{Peeples} et~al.}{2014}]{peeples2014}
{Peeples} M.~S.,  {Werk} J.~K.,  {Tumlinson} J.,  {Oppenheimer} B.~D.,
  {Prochaska} J.~X.,  {Katz} N.,   {Weinberg} D.~H.,  2014, \mn@doi [\apj]
  {10.1088/0004-637X/786/1/54}, \href
  {https://ui.adsabs.harvard.edu/abs/2014ApJ...786...54P} {786, 54}

\bibitem[\protect\citeauthoryear{{Priestley}, {Barlow}  \& {De
  Looze}}{{Priestley} et~al.}{2019}]{priestley2019}
{Priestley} F.~D.,  {Barlow} M.~J.,   {De Looze} I.,  2019, \mn@doi [\mnras]
  {10.1093/mnras/stz414}, \href
  {https://ui.adsabs.harvard.edu/abs/2019MNRAS.485..440P} {485, 440}

\bibitem[\protect\citeauthoryear{{Shetty}, {Kauffmann}, {Schnee}, {Goodman}  \&
  {Ercolano}}{{Shetty} et~al.}{2009}]{shetty2009b}
{Shetty} R.,  {Kauffmann} J.,  {Schnee} S.,  {Goodman} A.~A.,   {Ercolano} B.,
  2009, \mn@doi [\apj] {10.1088/0004-637X/696/2/2234}, \href
  {https://ui.adsabs.harvard.edu/abs/2009ApJ...696.2234S} {696, 2234}

\bibitem[\protect\citeauthoryear{{Whitworth} et~al.,}{{Whitworth}
  et~al.}{2019}]{Whitworthetal2019}
{Whitworth} A.~P.,  et~al., 2019, \mn@doi [\mnras] {10.1093/mnras/stz2166},
  \href {https://ui.adsabs.harvard.edu/abs/2019MNRAS.489.5436W} {489, 5436}

\bibitem[\protect\citeauthoryear{{Williams}, {Baes}, {De Looze}, {Rela{\~n}o},
  {Smith}, {Verstocken}  \& {Viaene}}{{Williams} et~al.}{2019}]{williams2019}
{Williams} T.~G.,  {Baes} M.,  {De Looze} I.,  {Rela{\~n}o} M.,  {Smith} M.
  W.~L.,  {Verstocken} S.,   {Viaene} S.,  2019, \mn@doi [\mnras]
  {10.1093/mnras/stz1441}, \href
  {https://ui.adsabs.harvard.edu/abs/2019MNRAS.487.2753W} {487, 2753}

\bibitem[\protect\citeauthoryear{{Ysard}, {Jones}, {Demyk}, {Bout{\'e}raon}  \&
  {Koehler}}{{Ysard} et~al.}{2018}]{ysard2018}
{Ysard} N.,  {Jones} A.~P.,  {Demyk} K.,  {Bout{\'e}raon} T.,   {Koehler} M.,
  2018, \mn@doi [\aap] {10.1051/0004-6361/201833386}, \href
  {https://ui.adsabs.harvard.edu/abs/2018A&A...617A.124Y} {617, A124}

\bibitem[\protect\citeauthoryear{{Zubko}, {Mennella}, {Colangeli}  \&
  {Bussoletti}}{{Zubko} et~al.}{1996}]{zubko1996}
{Zubko} V.~G.,  {Mennella} V.,  {Colangeli} L.,   {Bussoletti} E.,  1996,
  \mn@doi [\mnras] {10.1093/mnras/282.4.1321}, \href
  {http://adsabs.harvard.edu/abs/1996MNRAS.282.1321Z} {282, 1321}

\makeatother
\end{thebibliography}




\appendix

\section{Non-constant dust optical properties}
\label{sec:change}

{ While we have shown that an apparent anticorrelation between $\kfh$ and $\sism$ can be produced by a constant dust opacity, the trend predicted by dust evolution models is for $\kfh$ to increase with increasing $\sism$. The exact observed trend produced by a given dust model will depend not only on the optical constants, but also the relation between $\sism$ and the volume density of the gas, which is the relevant quantity for dust evolution (among others). This is beyond the scope of this paper, but we can investigate the likely effect by increasing the value of $\kfh$ for the attenuated dust component, to represent grain growth in the denser gas.}

{ We do this by changing the minimum and maximum sizes of our power law grain size distribution to $10$ and $20 \um$ respectively, which increases the opacity to $\kfh = 16.8 \csg$. This results in a flat relation between the inferred $\kfh$ and $\sism$ for a given $\fext$, since the temperature of the attenuated grains is lower, and so the flux is entirely dominated by the unattenuated component. A more moderate increase in the grain size distribution to between $0.1$ and $1 \um$ for the attenuated component results in a much smaller increase in $\kfh$, and a negligible change to the results in Figure \ref{fig:ks}. Whether the results of \citet{clark2019} are consistent with an increasing dust opacity with gas density therefore depends on the details of the dust evolution model under consideration.}

\vspace{2.cm}
\bsp	
\label{lastpage}
\end{document}